\documentclass[preprint]{aastex}
\usepackage{graphicx}

\def\simgreater{\buildrel > \over \sim}

\def\app{$\approx$~}

\def\etal{et~al.\ }

\def\hub{\ifmmode H_\circ\else H$_\circ$\fi}
\def\kms{~km~s$^{-1}$\ }

\begin {document}

\title{A New Definition for the Ca4227 Feature: Is Calcium Really Underabundant in Early-type Galaxies?}

\author{L. C. Prochaska} 
\affil{Department of Physics and Astronomy, CB 3255, University of 
   North Carolina, Chapel Hill, NC 27599}
\affil{Electronic mail: chaska@physics.unc.edu}
\author{James A. Rose}
\affil{ Department of Physics and Astronomy, CB 3255, University of 
   North Carolina, Chapel Hill, NC 27599}
\affil{Electronic mail: jim@physics.unc.edu}
\author{Ricardo  P. Schiavon}
\affil{Department of Astronomy, University of Virginia,
P.O. Box 3818, Charlottesville, VA 22903-0818}
\affil{Electronic mail: ripisc@virginia.edu}

\begin{abstract}
We have investigated the abundance of calcium in early-type galaxies by measuring the strength of the Ca I 4227 absorption line in their integrated spectra. The database used is the large sample of early-type galaxy integrated spectra in Caldwell, Rose, \& Concannon (2003). We have measured Ca abundances from the Ca I 4227 feature both by using the Lick Ca4227 index and also by defining a new index, Ca4227$_r$, that avoids the CN4216 molecular band in the continuum on the blueward side of the line. With the new index definition we measure Ca abundances that are systematically $\sim$0.3 dex higher than with the Lick Ca4227 index. The result is that with the new index definition we obtain higher [Ca/Fe] abundances in early-type galaxies which are more consistent with their well known [Mg/Fe] over-abundances. Hence, we suggest that Ca might be slightly enhanced, relative to Fe, in early-type galaxies.

\end{abstract}

\keywords{galaxies: abundances --- galaxies: evolution --- galaxies: stellar content}

\section{Introduction}

Accurate determinations of the chemical abundances in galaxies, based on
observations of their integrated spectra and on population synthesis
models, provide a fundamental contribution to deciphering the evolution of
galaxies.  While initial work focused primarily on determining an overall
heavy element abundance in galaxies, for some time now it has been recognized
that the chemical abundance patterns of other galaxies can be distinctly non-solar.  The most
striking example of non-solar abundance ratios is evident in the enhanced
[Mg/Fe]$\sim$+0.3 observed in the more massive early-type galaxies (Peletier 1989, Worthey \etal 1992, Davies \etal 1993, Kuntschner 2000, and references therein).
It is generally concluded that the enhanced magnesium-to-iron abundance
implies an overall enhancement of $\alpha$ elements relative to iron.  
Enhanced [$\alpha$/Fe] could result from a rapid star formation and chemical
enrichment history in massive early-type galaxies, when compared to the
$\sim$5 Gyr enrichment history of the pre-solar nebula (Worthey \etal 1992, Matteucci 1994, Thomas \etal 1999, Kuntschner 2000, and references therein).  A
counterpart to the $\alpha$-element enhancement in galaxies is found in
Galactic globular clusters, for which similar non-solar [$\alpha$/Fe] is
seen, and for which rapid enrichment histories are expected, given their very old ages.

Given the scenarios discussed above, it has come as a surprise that
several recent studies of the calcium abundances in early-type galaxies 
indicate that calcium does not track its $\alpha$-element companion Mg, but
rather that [Ca/Mg]$\sim$-0.3, and [Ca/Fe]$\sim$0.0 (Vazdekis \etal 1997, Worthey 1998, Trager \etal 1998, Henry \& Worthey 1999, Thomas \etal 2003, and references therein).  In
addition, Ca line strengths show a weak correlation with galaxy velocity
dispersion (thus galaxy mass), while Mg, as measured by the Lick Mg~$b$ index, exhibits a
steep trend with $\sigma$.  In short, unlike other $\alpha$ elements, calcium abundances seem to track those of iron in giant early-type galaxies. Thomas \etal (2003) have investigated several explanations for how
Ca can be decoupled from other $\alpha$ elements.  Except for the scenario in
which Ca is preferentially depleted onto dust grains, these scenarios all
require rather extreme conditions in terms of the initial mass function and/or
star formation rates, thus implying important modifications to previously
assumed star formation histories in galaxies. 

The situation concerning calcium in galaxies is rendered even more 
puzzling by the fact that little evidence exists for an
underabundance of calcium with respect to other $\alpha$ elements in our
Galaxy, even among the moderately metal-poor field stars in the solar 
neighborhood and thick disk stars which exhibit strong non-solar abundance 
ratio trends (Wheeler \etal 1989; Edvarsson \etal 1993; Reddy \etal 2003). 
Some evidence has been reported for calcium (and titanium) underabundances 
relative to the other $\alpha$ elements in Galactic bulge giants (McWilliam
\& Rich 1994).  Furthermore, Zoccali \etal (2004) report low [Ca/$\alpha$]
in giants in the metal-rich globular cluster NGC 6528, but both 
Carreta \etal (2001) and Origlia, Valenti, \& Rich (2005) find that Ca tracks
the other $\alpha$ elements in this cluster.  Discrepancies among these
studies undoubtedly result from the difficulty in conducting reliable high
dispersion abundance analyses for relatively cool, low-gravity
stars in regions of high and variable reddening. 

Clearly, the
abundance of calcium potentially provides a crucial insight
into their chemical enrichment histories. 
The low calcium abundances estimated in galaxies come from measurements of three different
spectral features.  The first is the Ca I 4227 resonance line, which has 
been measured in the Lick spectral index system, as defined in Worthey \etal (1994),
and then redefined in Trager \etal (1998).  A second feature is the Ca 4455
line, which has been measured in integrated spectra of early-type galaxies by, e.g., Trager \etal (1998) and Vazdekis \etal (1997). Third, there is the Ca
triplet at \app 8500\AA, which has been extensively studied in stars, globular
clusters, and early-type galaxies (Cohen 1979, Faber and French 1980, Alloin and Bica 1989, Idiart \etal 1997, Schiavon, Barbuy, \& Bruzual (2000),  Cenarro \etal 2001, Vazdekis \etal 2003).  Each of these spectral 
features has associated difficulties in regard to obtaining a clean
abundance measurement of calcium.  The Ca4455 feature is a blend with Fe
(Tripicco and Bell 1995), which, as noted above, is known to be underabundant relative to 
$\alpha$ elements in early-type galaxies.  The Ca triplet in the red has
three associated problems.  The first is that while measurements of the 
triplet correlate well with metallicity in metal-poor systems, little
abundance sensitivity is seen at higher metallicities (Alloin and Bica 1989, Vazdekis \etal 2003, and references therein).
Second, because the feature is located far in the red, the stars dominating
the integrated light of galaxies at the triplet come from the upper part of the
red giant branch (RGB) and the asymptotic giant branch (AGB). Due to the uncertainties in the modelling of the atmospheres of such cool giant stars, their elemental abundances are highly uncertain. In fact, even the iron abundances of the M giant stars included in the stellar libraries underlying current stellar population synthesis models are poorly known, if not entirely unknown. In addition, the
mean temperature of the stars dominating the light at 8500 \AA \ will be 
highly dependent on accurately modeling the luminosity function along the
upper RGB and AGB, which is difficult to achieve given the vagaries of mass
loss and sporadic episodes of elevated nuclear burning.  Third, the Ca triplet
is potentially contaminated by TiO features and by the Paschen lines
(Alloin and Bica 1989, Cenarro \etal 2001), which has led to the definition of a CaT$^*$ index that is
meant to minimize those problems (Cenarro \etal 2001).  Finally, the Ca4227 line lies
in a crowded spectral region in the blue, making it difficult to define an
index that cleanly measures the abundance of Ca.  The situation is exacerbated
by the presence of the strong CN4216 band just blueward of the Ca4227 feature. To further underscore the ambiguities surrounding what the Ca II triplet
and the CaI4227 features actually measure, Cenarro \etal (2004) have reported
divergent behavior in Ca abundances extracted from the two features in field
versus Coma cluster galaxies.

In this paper we focus on the Ca4227 feature in the integrated spectra of
early-type galaxies.  In \S2 we discuss the sample of galaxy spectra and
population modeling procedures that form the basis of our study.  In \S3 we
discuss the difficulties in measuring the Ca4227 line and define a new
index that is designed to minimize the problems.  In \S4 we present the
results on Ca abundances in galaxies obtained with our new index, and show
how these results differ substantially from those obtained with the previous
Ca4227 index. In \S5 we present systematic tests of our new index. In \S6 the implications of our results are summarized.

\section{Observational Data and Models}

\subsection{Observations}

The observational data used in this paper consist of spectra of 175 early-type
galaxies obtained with the 1.5 m Tillinghast Telescope, FAST spectrograph, and
a Loral 512 x 2688 pixel CCD \citep{Fabricant98} at the F. L. Whipple
Observatory.  The spectra were obtained at a dispersion of 0.75 \AA/pixel and
a resolution of 3.1 \AA \ FWHM, and are further described in Caldwell, Rose, \& Concannon 2003 (hereafter CRC).  A key point is that the galaxy sample covers early-type
galaxies over an extensive range in velocity dispersion, $\sigma$, from 50 \kms to more
than 300 \kms. As in CRC we restrict ourselves to the galaxies in the sample with $\sigma$$\leq$230 \kms.

\subsection{Population Synthesis Models}

To extract useful information about the Ca abundances in galaxies, as well as
abundances of other $\alpha$-elements and of Fe, requires comparison of the
observed line strengths in galaxy spectra to the predictions of a grid of
stellar population synthesis models covering a range in age and chemical
composition.  In this paper we use the population synthesis models of
Schiavon (2005) which are fully described in that paper.  Briefly, the
population models rely on the Padova isochrones of Girardi \etal (2000) and Salasnich \etal (2000) and 
a new set of fitting functions to the spectral indices used in this paper.
Equivalent widths of spectral indices are measured from the spectral library of Jones (1999). Theoretical stellar parameters are combined with these equivalent widths and polynomial fitting functions are computed. Predictions of integrated line indices are then generated by combining these fitting functions with the Padova theoretical isochrones. 
Further details can be found in Schiavon (2005).

\section{A New Ca4227 Index}

\subsection{Problems in Defining a Ca4227 Index}

\begin{figure}
\plotone{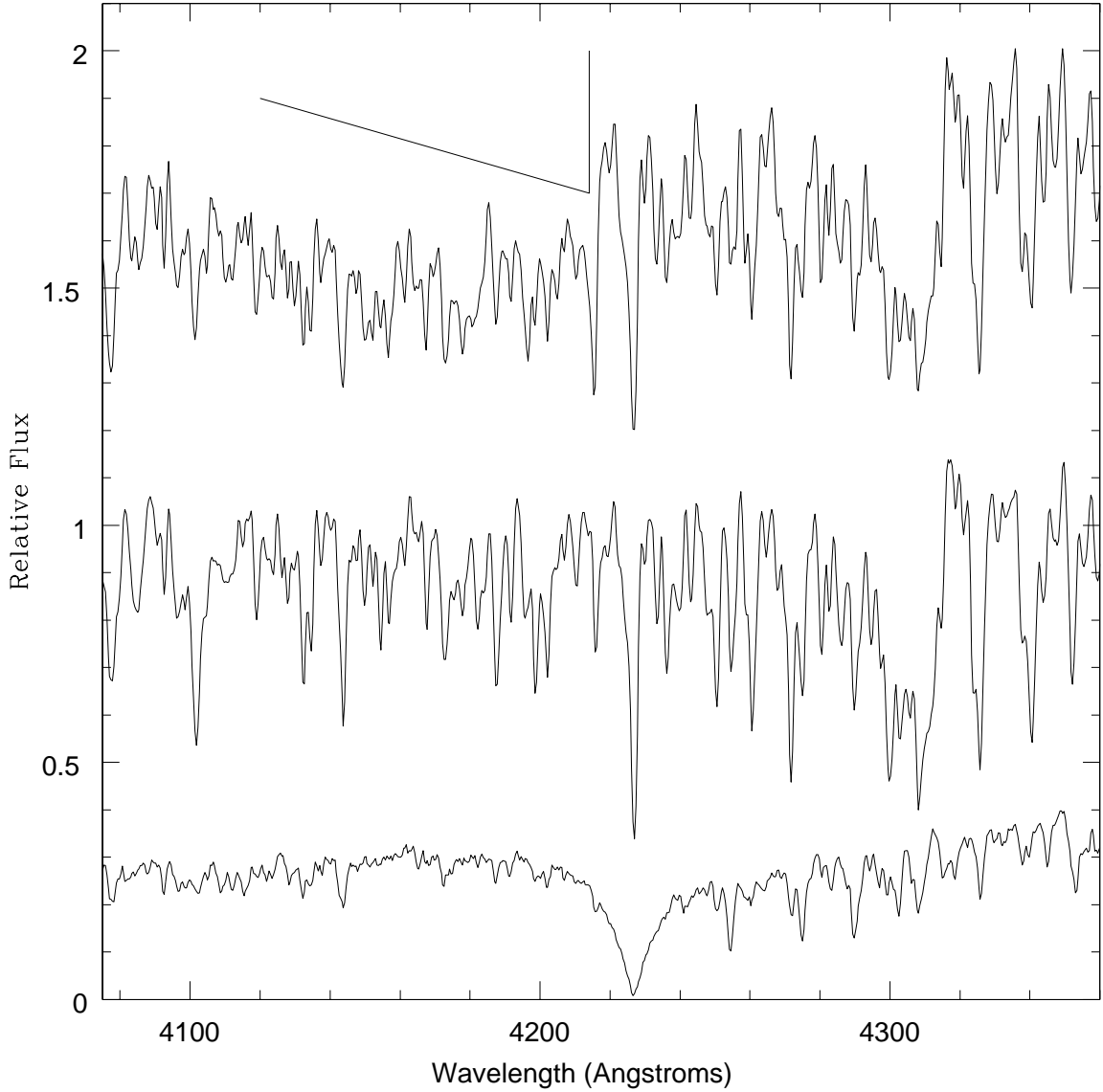}
\caption{Representative spectra of three stars in the region of the Ca I 4227
absorption line illustrate the difficulty in defining an index that cleanly
measures the feature.  The bottom spectrum is of the M dwarf G227-46, the 
middle spectrum is of the G5 dwarf HD115617, and the top spectrum is of the
K0III star HD115004.  The wedge at the top illustrates the position of the
CN4216 molecular band, with the bandhead at 4216 \AA.  All spectra have a 
resolution of $\sim$1.2 \AA \ FWHM.}
\label{fig:stellar}
\end{figure}

As mentioned in the Introduction, a critical issue for reliably measuring
element abundance ratios in the integrated spectra of galaxies is to
isolate the absorption feature of interest from other spectral features that 
may be overlapping in wavelength.  As well, since an accurate assessment of the
line strength requires the location of the stellar continuum on either side
of the absorption line, excessive contamination of the nearby continuum is in 
principle to be avoided.  The Ca4227 feature is in a particularly problematic
wavelength region from two considerations.  First, line crowding is especially
severe in the blue, hence it is difficult to isolate the Ca4227 line core and
its potentially extensive wings from other features in that region.  The result
is that a truly clean definition of Ca4227 is quite difficult to achieve.  The
problem of line crowding is illustrated in Fig.~\ref{fig:stellar}, 
where we have plotted the 
spectra of three representative stars.  The spectra have been taken from the
``Indo-US Library of Coude Feed Stellar Spectra'' (Valdes \etal 2004), which
consists of spectra of 1273 stars at a resolution of $\sim$1.2 \AA \ FWHM.
The spectrum on the bottom is that of an M dwarf, and illustrates the extensive wings
of Ca4227 in cool dwarfs.  Thus to obtain an assessment of the complete 
equivalent width of the line requires moving the neighboring continuum
bandpasses out by at least 30 \AA \ from the line center.  The middle and 
top spectra are respectively from a G5V and a KOIII star, which are the spectral types characteristic of the stars that dominate the integrated light of stellar populations in the blue spectral region. They indicate 
the degree of line crowding characteristic of stars that contribute to the 
integrated light of intermediate-age and old stellar population in the 
blue spectra region. Moreover, they illustrate a fundamental issue in 
defining the Ca4227 index.  Namely, at only 10 \AA \
blueward of the Ca4227 line center lies a strong feature which is due to a CN bandhead at $\lambda \sim$ 4216 \AA \. The CN4216 molecular band consists of an extensive series of features
that are most pronounced at the bandhead, but extend blueward for $\sim$100
\AA.  The CN4216 band, which is prominent in cool low-gravity stars,
is demarcated by the wedge at the top of Fig.~\ref{fig:stellar}. It is clearly not feasible to find anything like a true continuum around the Ca4227 line.

In principle, of course, it should not really matter whether a true continuum
is established for defining a spectral index, as long as the observed index in 
a galaxy integrated spectrum is compared to reliable stellar population models,
based on a stellar library that accurately reflects the spectra of the 
constituent stars in the galaxy.  In practice, however, the stellar abundance
patterns in early-type galaxies (at least in the more massive ones), appear
to be distinct from the abundance pattern in Solar Neighborhood stars, being
enhanced in the $\alpha$ elements such as Mg (Worthey \etal 1992, Davies \etal 1993, 
Kuntschner 2000, and references therein).  As shown by Trager \etal (1998), the strengths of CN bands in $\alpha$-enhanced, massive early-type galaxies tend to be stronger than predicted by the models, probably due to an overabundance ofnitrogen (Worthey 1998). Therefore, the 
strength of the CN4216 band may be underestimated in population synthesis models
for the integrated spectrum of an early-type galaxy, due to the use of a
spectral library based on Solar Neighborhood stars.  The result will be an
overestimate of the predicted strength of Ca4227 at a given ``metallicity'', due
to an underestimate of the contamination from CN in the continuum bandpass to 
the blue of Ca4227.  Consequently, the derived Ca abundance in the galaxy will
be underestimated.

\subsection{Definition of the Ca4227$_r$ Index}

\begin{figure}
\plotone{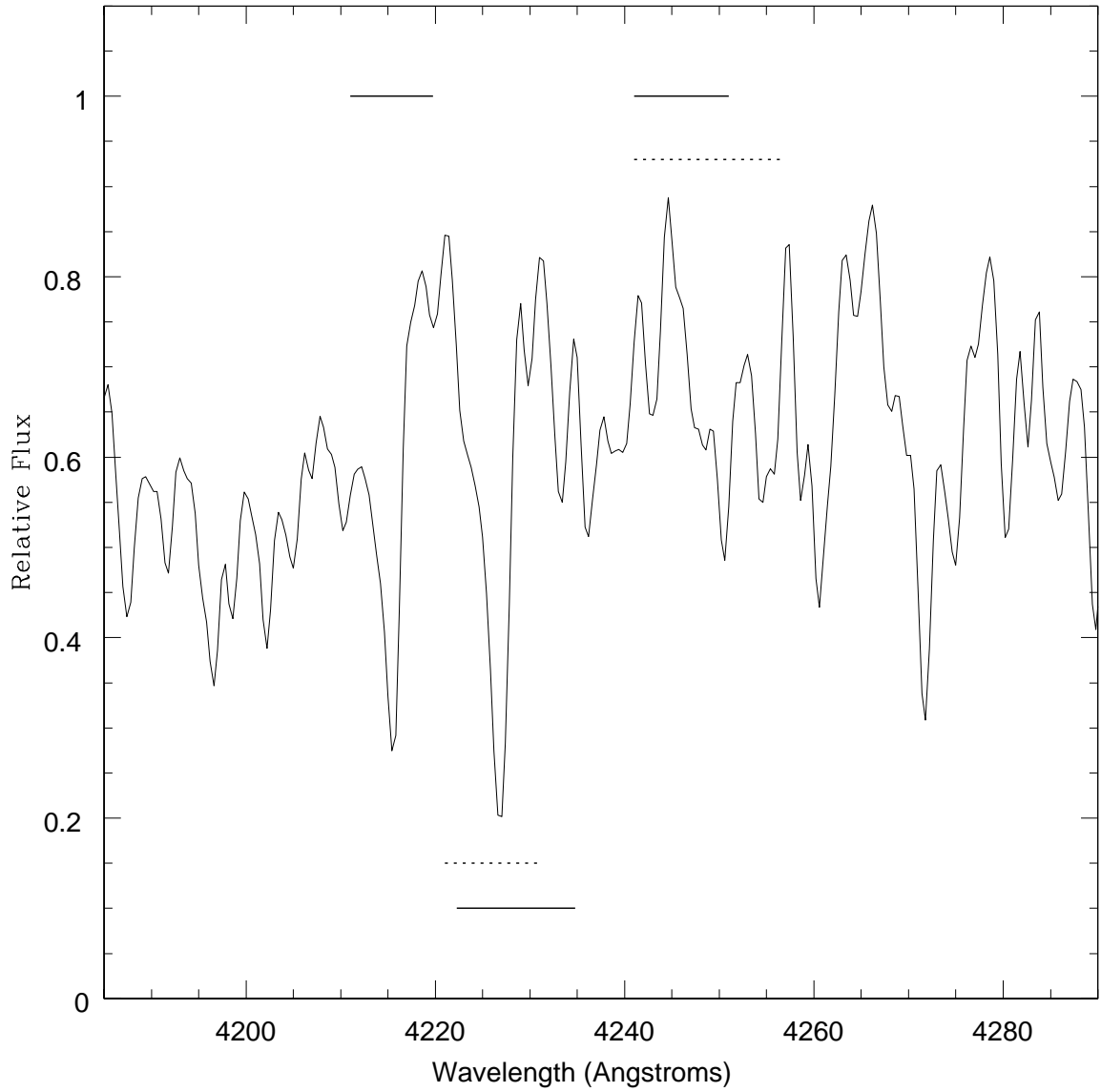}
\caption{The spectrum of the same K0III star, HD115004, as in 
Fig.~\ref{fig:stellar} is plotted in the vicinity of the Ca I 4227 absorption
line, with dotted lines indicating the central bandpass and the red continuum
bandpass used in our Ca4227$_r$ index. For comparison, the solid lines denote
the central bandpass and blue and red continuum bandpasses of the Lick
Ca4227 index.}
\label{fig:bands}
\end{figure}

From the discussion above, it should  be evident that two issues need special
consideration in defining an index that best isolates the Ca4227 absorption 
line.  The first is that the generally high degree of line crowding in the
region of Ca4227 makes it advantageous to use the narrowest possible bandpass in
isolating Ca4227 from other contaminating features.  On the other hand, we
must take into account that the line wings can be quite extensive, as 
illustrated by the M dwarf in Fig.~\ref{fig:stellar}.  However, from the standpoint of galaxy
(and globular cluster) integrated spectra, M dwarfs contribute little to the
integrated light, hence the necessity for measuring the wings of Ca4227 is
greatly overstated from the M dwarf spectrum.  Regardless, there is also the danger
in too narrowly confining the Ca4227 measurement to only the central few \AA \
such that the results will then be highly sensitive to velocity broadening effects
in the galaxy spectrum, which redistribute flux from the line
core into the wings. Moreover, very narrow index passbands require very high S/N spectra in order to be measured accurately.  While in principle it is possible to accurately measure the velocity
dispersion of the galaxy, as well as the intrinsic resolution of the spectrum,
an index highly dependent on resolution effects is best avoided.  As a 
compromise, we have chosen to measure a bandpass centered on Ca4227 that is
illustrated by the dotted line in Fig.~\ref{fig:bands}.  There we have 
replotted the spectrum
of the same K0III star from Fig.~\ref{fig:stellar}, and also, for reference, 
plot the central bandpass of the Lick Ca4227 index as a solid line. 
Note that our bandpass is
similar to the Lick bandpass, with a slight shift in central wavelength.  

The second consideration regards the placement of the pseudo-continuum bandpasses.
The Lick Ca4227 blue and red continuum bandpasses are represented by solid lines
at the top of Fig.~\ref{fig:bands}.  Note that although the blue continuum bandpass is quite
narrowly defined, and situated close to the Ca4227 feature, it unavoidably
falls into the CN4216 molecular band.  The only way to avoid the CN band would
be to define a very narrow blue continuum bandpass that isolates the peak just
longward of the CN bandhead at 4216 \AA.  This method was followed in Rose 
(1994).  However, it is subject to the above-mentioned resolution effects, in
that increased velocity broadening will remove flux from the localized
pseudocontinuum peak.  Consequently, the approach we have chosen for the new
Ca4227$_r$ ({\em r} for {\em red} bandpass) index is to give up the idea of a
blue continuum bandpass altogether.  Instead, we have defined a red continuum
bandpass, as denoted by the dotted line redward of Ca4227, and assume a flat
continuum in the region of Ca4227 in calculating the pseudo-equivalent width.
In effect, the Ca4227$_r$ index is simply the mean flux ratio between line and 
continuum bandpasses, but conveniently recast in terms of an equivalent width.
The one disadvantage of this method is an increased sensitivity to poor flux
calibration of the spectra and/or reddening (see \S5).  The exact 
wavelength limits of
the Ca4227$_r$ and the Lick Ca4227 indices are given in Table ~\ref{tab:indexdef}.

\subsection{Measurement of Spectral Indices}\label{sect:meas}

In this paper we make use of a number of key spectral line indices that are
defined in the ``Lick'' system by Worthey \etal (1994).  In particular, we 
use the Lick H$\beta$, Mg~$b$, Fe4383, and Ca4227 indices, in addition to our
newly defined Ca4227$_r$ index.  For the Lick indices we use the bandpass
definitions given in Worthey \etal (1994).  In the true Lick system, as it
originated with the IDS detector, the spectral resolution varies with
wavelength (as specified in Worthey \& Ottaviani 1997), and the spectra are on the 
instrumental response of the IDS.
On the other hand, the CRC03 galaxy spectra used in this study are 
flux-calibrated, and are
taken at a spectral resolution of 3.1 \AA \ FWHM and then all spectra are
smoothed to a common velocity broadening, $\sigma$, of 230 \kms (see CRC03
for details).  It turns out that this effective resolution of the galaxy
spectra, at both H$\beta$ and Mg~$b$, almost precisely matches the
resolution of the Lick system at those indices.  To place ourselves closer to
the Lick system, before calculating the Fe4383 and Ca4227 indices we smoothed
the galaxy spectra by an extra amount to attain the same resolution as the
Lick system.  Note, however, that our results are not exactly on the true
Lick system, since the galaxy spectra are flux-calibrated, while the Lick
system, as mentioned above, is based on the IDS instrumental response.  Thus
although we refer hereafter to the ``Lick'' Ca4227, H$\beta$, Mg~$b$, and
Fe4383 indices throughout the paper, the reader should note that we are not
quite on the standard system.  Due to the fact that the Lick indices always
involve continuum bandpasses on both sides of the absorption feature of 
interest, the nature of the instrumental response has a very weak effect on 
the measured index.  Note also that the models from Schiavon (2005) are also
based on fitting functions from a flux-calibrated stellar spectral library.
Thus although the library spectra have been smoothed by the appropriate amount
for each Lick index, as in the case of our galaxy spectra, they are based on
flux-calibrated stellar data.  For the
newly defined Ca4227$_r$ index, no additional smoothing has been carried out
beyond getting all galaxy spectra onto a common velocity broadening of
230 \kms, in addition to the intrinsic resolution of the spectra of 3.1 \AA \
FWHM. We  note that the model predictions by Schiavon (2005) for the Ca4227$_r$ index were computed for the exact same broadening as the data discussed here.

Finally, all of the indices, with the exception of H$\beta$, are measured in the galaxy spectra using
a modified version of the automated LECTOR program made publicly available by
Alexander Vazdekis\footnote{http://www.iac.es/galeria/vazdekis/vazdekis\_models.html}.
 The H$\beta$ indices are from CRC data which have been corrected for both the effects of emission and non solar abundance ratios. These correction procedures are documented in CRC. The measured Ca4227$_r$ and Lick Ca4227 index values for the spectral library of Jones (1999) used in the models are given in Table ~\ref{tab:jonesindices}.
The measured Ca4227$_r$ and Lick Ca4227 index values for each galaxy in the CRC sample along with the $\pm$1$\sigma$ uncertainties in the indices are given in Table ~\ref{tab:galaxyindices}.  Uncertainties in
the Ca4227$_r$ indices were determined from the r.m.s. scatter in multiple
exposures for many galaxies (see CRC for details).  Uncertainties in the
Lick spectral indices used by us are in principle already determined by CRC.
However, since we have recalculated the Fe4383 and Lick Ca4227 indices after
smoothing to the Lick resolution specified in Worthey \& Ottaviani (1997), we
determined new 1$\sigma$ uncertainties by calculating the r.m.s. scatters in
the Lick indices in multiple exposures for a subsample of the CRC galaxies.
From this subsample we determined an overall correction factor from the
published CRC uncertainties to those in our index determinations carried out
at the Lick resolution. In the case of 47 Tuc, where multiple exposures were not available, we calculated the uncertainties from the signal to noise ratio per pixel integrated over the specified bandpass.  The characteristic $\pm$1$\sigma$
error bars shown in Fig.~\ref{fig:trend} and all subsequent figures represent 
the mean errors for the entire CRC sample.

\section{Results}\label{sect:resuls}

The primary goal of this study is to evaluate the behavior of [Ca/Fe]
and [Ca/$\alpha$] in galaxies.  As was discussed in \S3, there is reason to
suspect that previous attempts to measure the calcium abundance through the
Ca I $\lambda$4227 feature may have been compromised by CN contamination in
the blue continuum bandpass.  Consequently, we have defined a new index,
Ca4227$_r$, that is designed to remove the CN contamination problem.  In this
section we compare the behavior of the new calcium index with that of the
Lick Ca4227 index.

\subsection{Empirical Trends}\label{sect:empirical}

\begin{figure}
\plotone{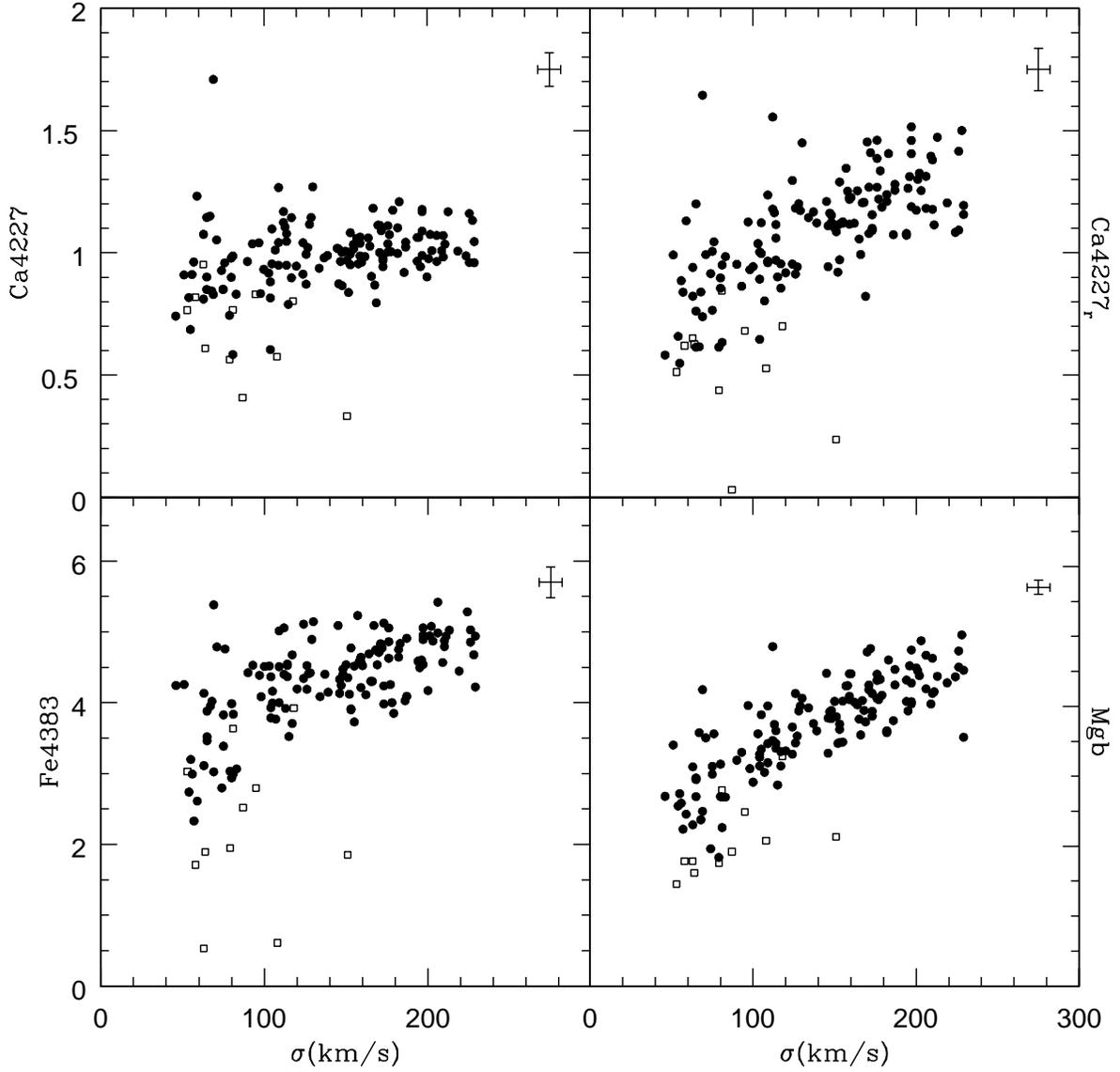}
\caption{Empirical trends of Ca, Mg, and Fe versus galaxy velocity dispersion
are presented for the CRC integrated spectra of early-type galaxies. The open squares denote young galaxies (EWH$\beta$ $\simgreater$ 3.0).
In the top left panel are plotted data for the Lick Ca4227 index, 
while that for our newly defined Ca4227$_r$ index is shown in the top right
panel.  For reference, the trends for the Mg~$b$ and Fe4383 indices are plotted
in the lower left and right panels respectively. Error bars represent the mean r.m.s. errors among repeated galaxy observations.}
\label{fig:trend}
\end{figure}

We begin by examining the behavior of the old and new Ca indices in a
strictly empirical sense, i.e., without reference to population synthesis
modeling.  Specifically, we use the large database of early-type galaxy
spectra from CRC.  One attribute of the CRC dataset is that it covers a
large range in galaxy velocity dispersion (hence mass), thus we can 
examine trends in Ca versus other elements over a large mass range.
Several major studies have established that the abundance of Mg increases
substantially with galaxy velocity dispersion, $\sigma$, while the increase 
in Fe with $\sigma$ is 
less striking (Jorgensen 1997, Kuntschner 2000, and references therein).  
We reproduce the behavior of Mg and Fe with
$\sigma$ found by CRC in their dataset in the bottom left and right
panels of Fig.~\ref{fig:trend}.  We have used the Mg~$b$ and Fe4383
indices to track the abundances of Mg and Fe, respectively.  There is a
well-defined slope to the Mg~$b$ versus $\sigma$ relation that extends
from the lowest to the highest-$\sigma$ galaxies.  The only exceptions
are those galaxies dominated by young stellar populations especially prominent 
at the low $\sigma$ end, which scatter to low Mg~$b$ values (to separate these young galaxies from the rest of the sample, we have denoted by open squares galaxies with EW H$\beta$ $\simgreater$ 3.0).  In contrast,
the trend in Fe4383 versus $\sigma$ is not nearly so steep, and appears to
flatten out at the high-$\sigma$ end.  Again, there is a scatter of
younger galaxies to low Fe4383, especially at the low $\sigma$ end.  These
trends are extensively discussed in CRC.

In the top left and right panels of Fig.~\ref{fig:trend} we illustrate the 
behavior of both the Lick
Ca4227 index and our newly defined Ca4227$_r$ index.  There is a 
noticeable difference in the trends with $\sigma$ for the two indices, in that
the Lick Ca4227 index exhibits only a weak trend with $\sigma$, similar to that of the
Fe4383 index, while Ca4227$_r$ exhibits a strong trend, similar to that seen
in Mg~$b$.  This striking difference in the empirical behavior of the two
calcium indices suggests that indeed what one measures in the Ca4227 feature
is critically dependent on the measurement strategy.  However, to better
assess how the actual extracted abundance of calcium depends on the index
definition requires the use of population synthesis modeling, hence we
turn to that approach in what follows.

\subsection{Results from Population Synthesis}

\begin{figure}
\plotone{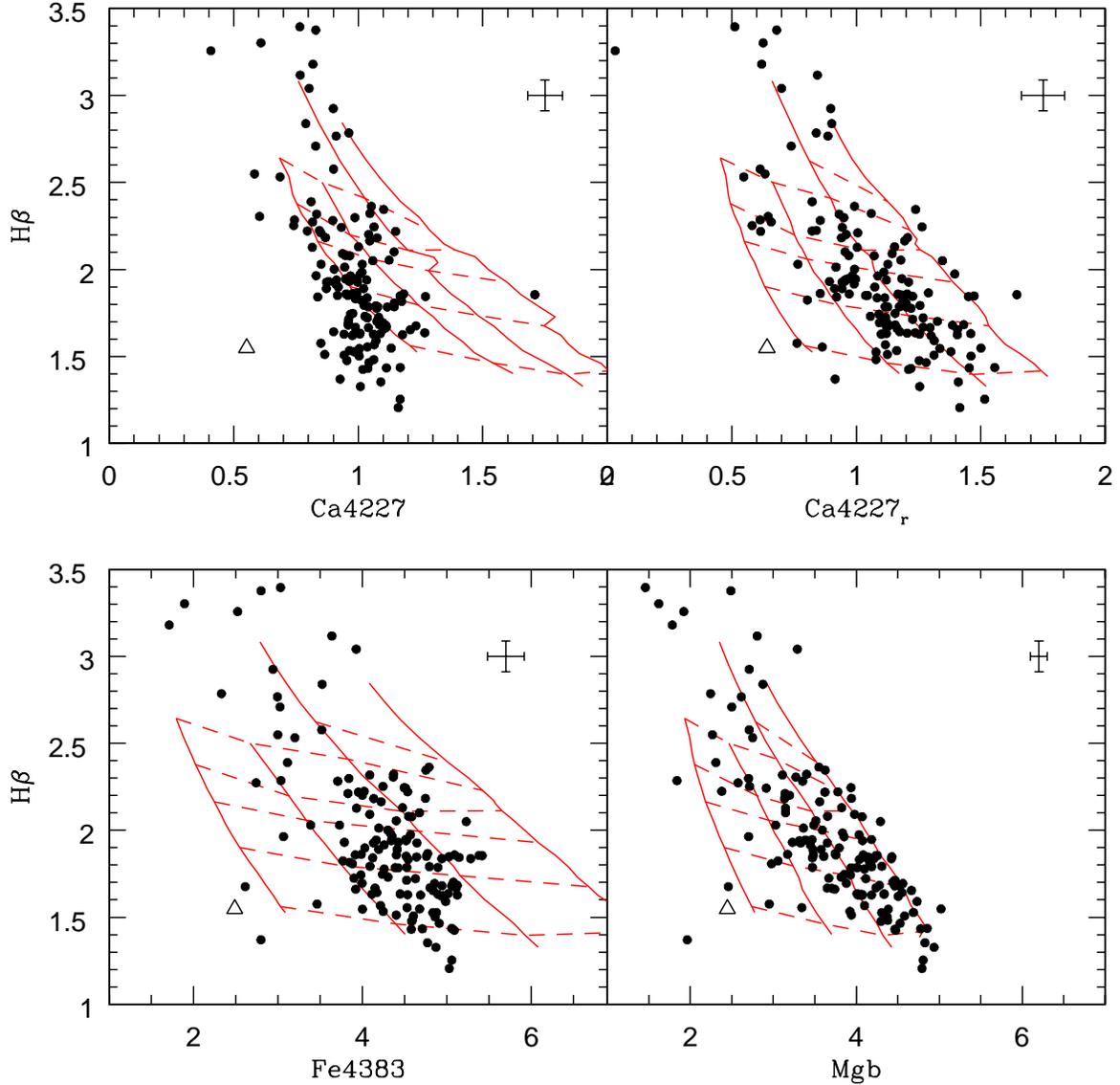}
\caption{Lick Ca4227, Ca4227$_r$, Mg~$b$, and Fe4383 versus H $\beta$ for the same CRC galaxy data 
as in Fig~\ref{fig:trend}. Also plotted is a data point for the
globular cluster 47 Tuc (open triangle).  Overplotted on the data points are 
the model grid lines of constant age and [Fe/H] from Schiavon (2005). The 
specific metallicities (solid lines) are, from left to right, 
[Fe/H]=-0.8, -0.4, 0.0, and +0.3.  The ages (dashed lines) are, top to bottom, 
2.0, 2.5, 3.5, 5.0, 7.9, and 14.1 Gyr. Error bars represent the mean r.m.s. errors among repeated galaxy observations.}
\label{fig:CaHbeta}
\end{figure}

The most widely used strategy for separating age from metallicity effects in
the integrated spectra of galaxies is to plot a primarily age-sensitive
Balmer line versus a metal-sensitive feature, and compare galaxy data to
models of single stellar populations in these plots.  Since the Lick
H$\beta$ index represents the most extensively used Balmer line strength 
indicator, we plot our calcium indices versus this spectral feature in
Fig.~\ref{fig:CaHbeta}.  The left hand panel features the Lick Ca4227 index, while the
right hand panel features our new Ca4227$_r$ index. For comparison, in the bottom left and right panels are plotted Fe4383 and Mgb, respectively, versus H$\beta$. In all panels the model
population grids come from the Schiavon (2005) models described in \S2.2,
while the galaxy data are from the CRC spectra.  The H$\beta$ indices in the
CRC data have been corrected both for the effects of emission and non solar abundance ratios.  These
correction procedures are documented in CRC. In the top left panel we
basically reproduce the result of, e.g., Thomas \etal (2003), that the 
abundance of calcium appears to be rather low, especially in comparison with that of magnesium.  While [Mg/H] ranges from slightly sub-solar in
low $\sigma$ galaxies to as much as +0.5 in the higher $\sigma$ galaxies, the
Lick Ca4227 index indicates that [Ca/H] is decidedly subsolar in most
early-type galaxies.  On the other hand, the Ca4227$_r$ plot indicates a
substantially higher mean [Ca/H] for early-type galaxies.  Specifically, the
mean [Ca/H] is now somewhat higher than solar.  Zeropoint uncertainties abound
in population synthesis modeling, especially when comparing element
abundances derived from one set of models with another one.  Consequently,
we consider the shift in the mean [Ca/H] of $\sim$+0.3 from the Lick
Ca4227 index to the Ca4227$_r$ index to provide the more compelling case for
our assertion that the Lick Ca index definition may have led to an
underestimate of Ca abundances in galaxies.  However, there is also some
evidence that the new Ca4427$_r$ zeropoint supplies a better 
representation of Ca abundances.  Specifically, we have plotted as a large
unfilled triangle the Ca4227
and Ca4227$_r$ indices for an integrated spectrum of the core of 47 Tuc, in
which the spectrograph slit was trailed across the core diameter of the
cluster.  From high-dispersion spectroscopic abundance analyses of individual 
RGB stars, the [Fe/H]
value for 47 Tuc is determined to be $\sim$-0.7, while [Ca/Fe]
estimates range from 0.0 to $\sim$+0.2 (Carretta \etal 2004).  In the left hand panel 
the Ca abundance inferred for 47 Tuc is clearly far below the expected value
of [Ca/H]$\sim$-0.7 (or perhaps -0.5), while in the right hand (Ca4227$_r$)
panel, the inferred Ca abundance is much closer to the expected value from the
high dispersion spectroscopy of giants.  Note that we expect a major effect on
the Lick Ca4227-derived abundance, since roughly half the stars in the core of
47 Tuc are CN-strong (Norris \& Freeman 1979).  

\begin{figure}
\plotone{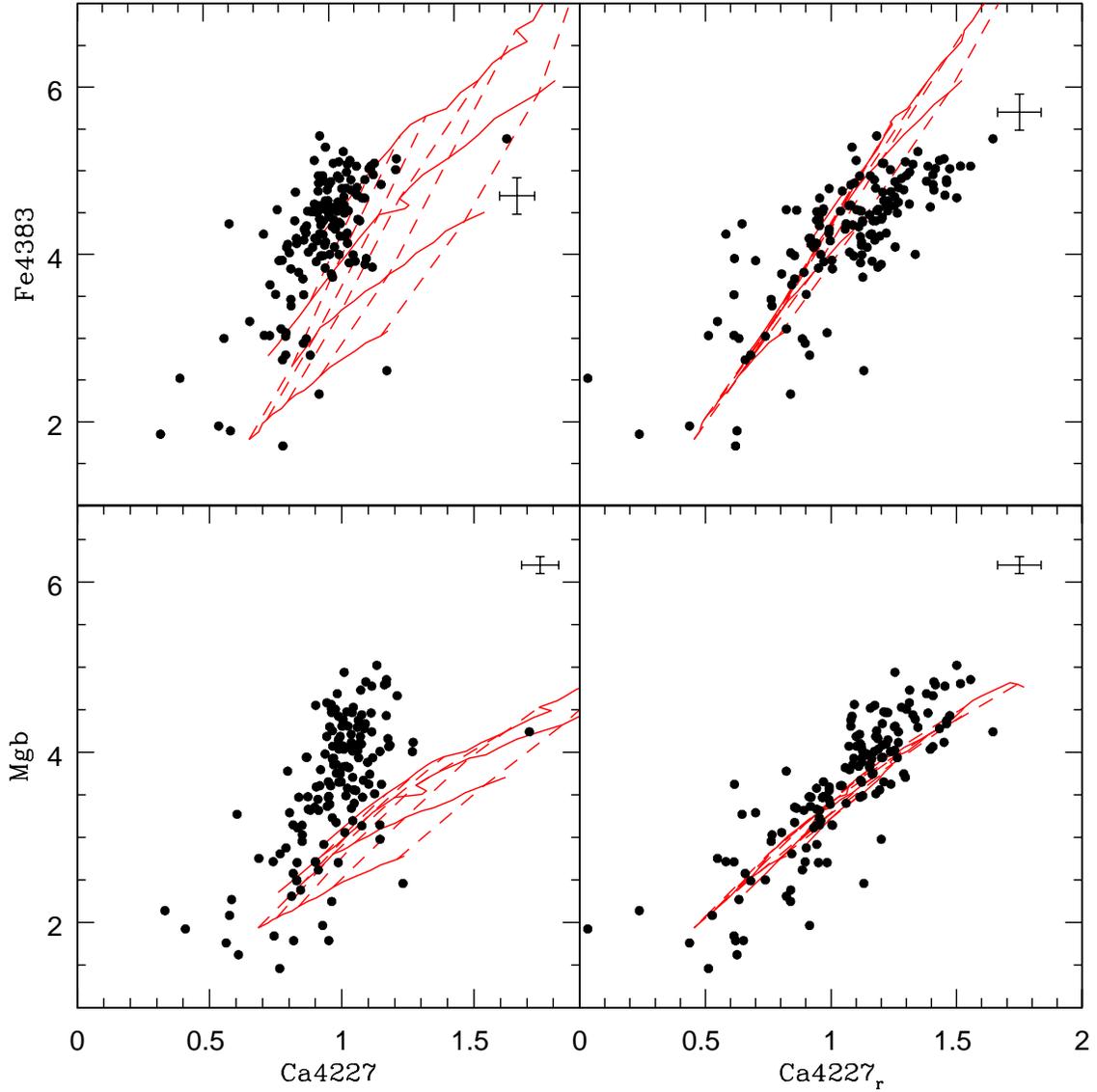}
\caption{Fe4383 (top panels) and Mg~$b$ (bottom panels) galaxy data from 
CRC is plotted versus the Lick Ca4227 (left panels) and the
new Ca4227$_r$ index (right panels).  The model grid lines are the same as
in Fig.~\ref{fig:CaHbeta}. In the top left and bottom left panels, lines of constant age are increasing from left to right and  metallicity increases from bottom up. Error bars represent the mean r.m.s. errors among repeated galaxy observations.}
\label{fig:NSAR}
\end{figure}

To further clarify the effect of Ca index definition on the derived Ca 
abundances, in Fig.~\ref{fig:NSAR} we plot the Lick Ca4227 index versus both 
Fe4383 (top 
left panel) and Mg~$b$ (bottom left panel) indices, while in the right hand
panels we plot Fe4383 and Mg~$b$ versus the new Ca4227$_r$ index.  The 
Schiavon (2005) model
grid lines are compared to the CRC early-type galaxy data in the four panels.
Since the models were computed under the assumption of solar abundance 
element ratios, not surprisingly, the model grid lines are highly degenerate
in Fig.~\ref{fig:NSAR}.  In the Lick Ca4227 plots, the CRC galaxy data points 
lie on top
of the grid lines in the Fe4383 diagram, whereas in the Mg~$b$ plot the
higher Mg~$b$ galaxies lie well separated from the grid.  Hence these two
panels reproduce the basic results that indicate an underabundance of Ca
relative to other $\alpha$ elements (as represented here by Mg) 
(Vazdekis \etal 1997, Worthey 1998, Trager \etal 1998, Thomas \etal 2003).
On the other hand, the opposite effect is seen in Ca4227$_r$, namely, that 
Ca is systematically slightly overabundant relative to Fe for the higher Ca4227$_r$
galaxies (top right panel) whereas it tracks Mg~$b$ very well (bottom right
panel).  

To summarize at this point, there is a systematic difference of $\sim$0.3
between Ca abundances derived from the Lick 4227 index versus the new
4227$_r$ index, in the sense that the Ca derived from the new index exceeds 
that from the old index.  Population synthesis modeling also indicates that
while the Ca abundance derived from the Lick index tracks that of Fe, the
Ca4227$_r$-derived Ca abundances closely track that of Mg, thereby 
indicating that with the new Ca index, Ca abundances appear to follow the trend
exhibited by other $\alpha$ elements.

\section{Systematics}

In addition to assessing how sensitive a spectral index is to the abundance of
the particular element being measured, it is important to determine how robust
the index is to realistic errors of measurement arising from uncertainties in
spectral resolution, wavelength zeropoint, and flux-calibration.

\subsection{Instrumental Resolution}
 We first consider to what extent
realistic uncertainties in spectral resolution may compromise results on the
abundance of Ca derived from the Ca4227$_r$ index.  Two effects must be
considered in this regard: uncertainties in determining the intrinsic 
resolution of the spectrograph/detector and uncertainties in determining the 
velocity dispersion of the particular galaxy.  

Regarding the intrinsic
spectral resolution, such information is typically generated from fitting
Gaussian profiles to individual lines in an arc spectrum while focusing the
spectrograph.  While normally the focusing procedure is optimized to produce
a uniform focus along the dispersion, it is not unusual to end up with up to
a 5\% variation in FWHM of the arc lines along the dispersion axis.  
Consequently, it is not necessarily valid to cite a single resolution for the
entire spectrum.  In addition, focus variations will occur during the night as
a result of temperature changes and flexure of the spectrograph during a long
exposure.  Hence we consider a 5\% error in the intrinsic resolution of the
equipment to be a reasonable figure.  

\begin{figure}
\plotone{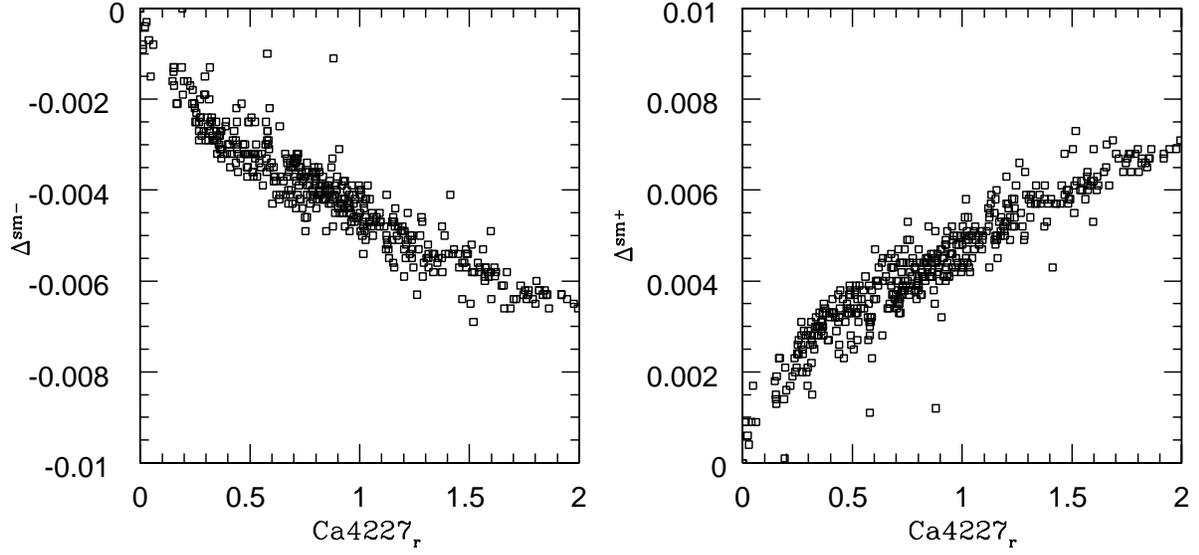}
\caption{The effect of spectral resolution on the measured Ca4227$_r$ index is
illustrated.  In the left panel we plot the difference between the index
measured at 3.1 \AA \ FWHM resolution with that at 2.95 \AA \ FWHM (in the
sense that $\Delta^{sm-}$ is the Ca4227$_r$ index at 3.1 \AA \ resolution minus
Ca4227$_r$ at 2.95 \AA \ resolution).  In the right panel, $\Delta^{sm+}$ is the
Ca4227$_r$ index at 3.1 \AA \ resolution minus Ca4227$_r$ at 3.25 \AA \ resolution.}
\label{fig:resn}
\end{figure}

To simulate the effect of intrinsic resolution effects we utilized stellar
spectra from the Indo-US Library of Coude Feed Stellar Spectra (Valdes \etal 
2004) mentioned earlier in \S3.1.  Since this spectral library is sampled at
much higher intrinsic spectral resolution (FWHM$\sim$1.2) than the galaxy 
spectra
from CRC (FWHM$\sim$3.1), any resolution variations in the stellar library will
be unimportant.  Furthermore, the high signal-to-noise ratio ($>$100:1 per
0.4 \AA \ pixel) minimizes scatter in the results due to photon noise in the 
data.  We Gaussian smoothed the stellar spectra to intrinsic resolutions of
2.95 \AA, 3.1 \AA, and 3.25 \AA \ FWHM, and then further broadened them to a
velocity dispersion of 230 \kms.  Thus the 3.1 \AA \ resolution spectra 
reproduce the resolution of our galaxy spectra (which are all smoothed as well
to a common velocity dispersion of 230 \kms) from which the Ca4227$_r$ index is measured, while the other two intrinsic
resolutions simulate a $\pm$5\% error in intrinsic resolution.  In 
Fig.~\ref{fig:resn} we
plot the resulting offset between the measured Ca4227$_r$ index of the spectra 
at a resolution of 3.1 \AA \ FWHM to spectra at 2.95 (left panel) and 3.25 \AA
\ (right panel).  This exercise has been carried out for stellar spectra
covering a wide range in spectral type.  As can be seen in Fig.~\ref{fig:resn}, 
systematic
shifts in the Ca4227$_r$ index are less than $\pm$0.006, which translates to
a negligible ($\sim$0.01 dex) effect on the derived Ca abundance from the index.

\subsection{Velocity Broadening}
\begin{figure}
\plotone{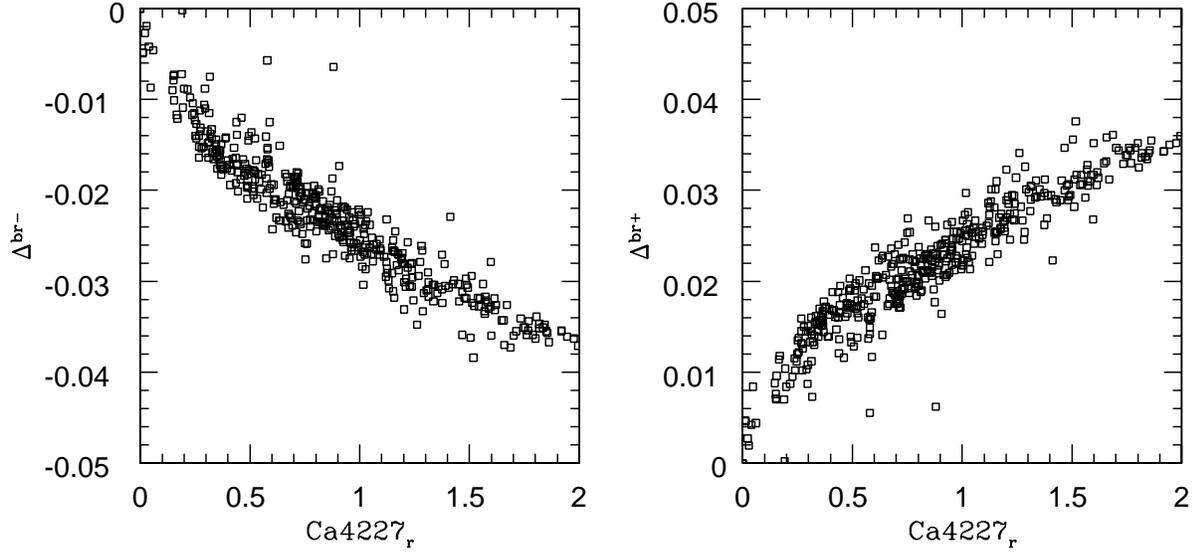}
\caption{The effect of Doppler broadening on the measured Ca4227$_r$ index is
illustrated.  In the left panel we plot the difference between the index
measured at a Doppler broadening, $\sigma$, of 230 \kms with that broadened
to 220 \kms (in the sense that  $\Delta^{br-}$ is the Ca4227$_r$ index at
$\sigma$ of 230 \kms minus Ca4227$_r$ at 220 \kms broadening).  In the right
panel, $\Delta^{br+}$ is the Ca4227$_r$ index at a $\sigma$ of 230 \kms minus
that at $\sigma$ of 240 \kms.}
\label{fig:broad}
\end{figure}

The observed spectral resolution is also degraded by velocity broadening of
spectral lines in galaxies.  Typical uncertainties in determining the
velocity dispersion, $\sigma$, are $\pm$10 \kms.  To simulate that effect,
we Gaussian broadened the same set of stellar spectra to velocity dispersions
of 220 \kms, 230 \kms, and 240 \kms (after first smoothing them to the
intrinsic spectral resolution of 3.1 \AA \ FWHM of the CRC galaxy data). In
Fig.~\ref{fig:broad} we show the systematic offset in the measured Ca4227$_r$ index that is caused by this amount of uncertainty in broadening.  Here the effects are larger,
but do not exceed $\pm$0.03.  As can be seen from the average spacing of the model grid
lines in Fig.~\ref{fig:CaHbeta}, an uncertainty of
$\pm$0.03 in the Ca4227$_r$ index translates into an uncertainty of $\pm$0.05
dex in the derived Ca abundance.

\begin{figure}
\plotone{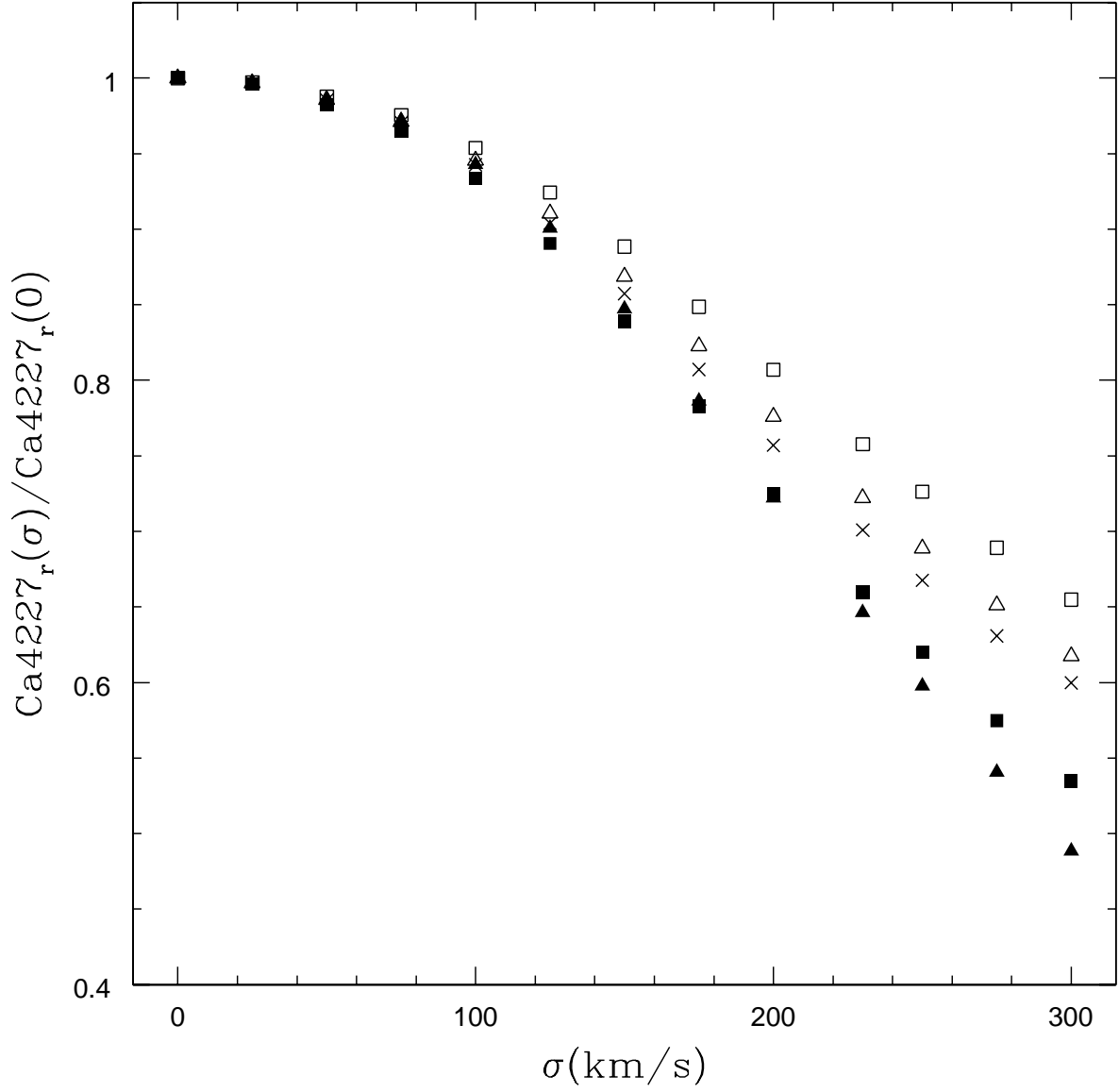}
\caption{The dependence of the new Ca4227$_r$ index on galaxy velocity dispersion is illustrated. The Ca4227$_r$ index, normalized to Ca4227$_r$ at $\sigma$ = 0\kms, is plotted against the velocity dispersion, $\sigma$. Results are plotted for 5 stellar spectra, covering a range in Ca4227$_r$ index. The stellar spectra are designated by open square, open triangle, cross, closed square, and closed triangle representing in increasing order a run in the Ca4227$_r$ index from 0.47 to 1.57.}
\label{fig:sig}
\end{figure}

To obtain a more comprehensive idea regarding the sensitivity of Ca4227$_r$ to velocity dispersion, we Gaussian broadened a set of 5 stellar spectra (covering a range in spectral type) from $\sigma$ = 0 to $\sigma$ = 300\kms. Ca4227$_r$ indices were measured for each and then normalized to the value of Ca4227$_r$ at $\sigma$ = 0\kms. Such normalized indices are shown in Fig.~\ref{fig:sig} with each spectrum designated by a separate symbol.

\subsection{Wavelength Zero Point}
\begin{figure}
\plotone{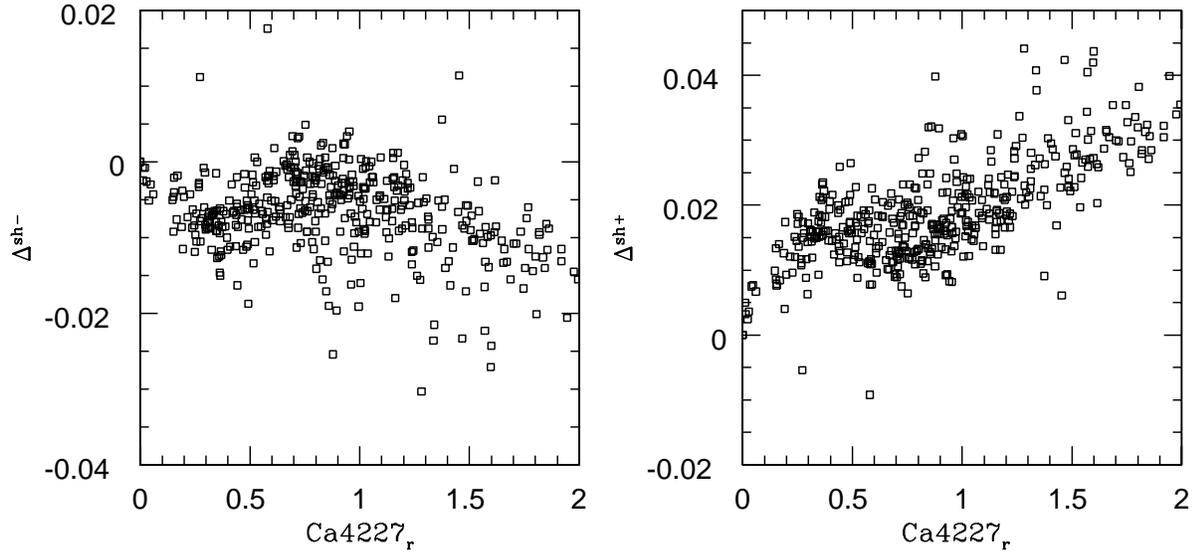}
\caption{The effect of a wavelength zero point shift on the measured Ca4227$_r$
index is illustrated.  In the left panel, the wavelength scale has been 
shifted by -0.5 \AA, and the difference in the measured Ca4227$_r$ index is in 
the sense of the unshifted index minus the shifted.  In the right hand panel
the same procedure has been carried out, except that that a shift of +0.5 \AA \
has been carried out.}
\label{fig:shift}
\end{figure}

We have simulated the effect of uncertainty in the wavelength zero
point of the spectra.  Since the signal-to-noise ratio of galaxy spectra
used for abundance determinations is generally high enough for a precise
redshift determination, we expect that the primary source of wavelength
zero point uncertainties will arise from spectrograph flexure.  A half pixel
uncertainty, which translates to 0.5 \AA \ at our spectral sampling, is
likely a generous allowance for flexure errors.  Accordingly, we have shifted
the same sample of stellar spectra by $\pm$0.5\AA, and the results are plotted
in Fig.~\ref{fig:shift}.  As in the case of velocity broadening, zero point shifts in
wavelength can introduce an uncertainty of $\pm$0.03 in the Ca4227$_r$ index,
or $\pm$0.05 dex in the derived Ca abundance.  To further test the effect of
wavelength zero point shift, we applied $\pm$0.5 \AA \ shifts to the CRC
galaxy spectra.  The results, while considerably noisier than for the very high
S/N ratio stellar spectra, are consistent with that derived for the stellar
spectra.

\subsection{Flux Calibration and Reddening Effects}

Finally, given that the new Ca4227$_r$ index was defined using only one continuum bandpass, it is important to be aware of the sensitivity to uncertainties in the slope of the continuum. Uncertainties in slope could arise from flux-calibration errors or reddening effects. In the spirit of the above analysis, we simulated the effect by assuming an error in continuum slope that translates into an error in B-V color of $\pm$ 5\%. A slope error of this magnitude results in an offset in the measured Ca4227$_r$ index of $\pm$0.01, which reflects a change in derived Ca abundance of  $\sim\pm$0.01.

 To summarize, realistic uncertainties in spectral resolution, wavelength zero point, and flux calibration lead to modest uncertainties in the Ca abundances derived from the Ca4227$_r$ index. The largest sources of uncertainty are wavelength zero point and velocity broadening which are at the level of $\pm$0.05 \ dex in the derived Ca abundance. Flux calibration and instrumental resolution are less problematic.

\section{Summary}

Our analysis of the Ca I 4227 absorption line in the integrated spectra of galaxies has led to the following conclusions.

(i) We have defined a new index, Ca4227$_r$, that includes only a single
continuum bandpass on the red side of the Ca feature at 4227\AA.  We do not
use a continuum bandpass blueward of the 4227 line due to the presence of
the CN4216 molecular band in that region.
The central bandpass for our newly defined Ca4227$_r$ index
is located between 4221 \AA \  and 4230.8 \AA; the red continuum bandpass is 
between 4241\AA \ and 4257 \AA.  

(ii) When applied to the integrated spectra of early-type galaxies, the new 
index produces a derived calcium abundance that is systematically greater 
than that determined from the Lick Ca4227 index of Worthey \etal (1994). 

(iii) While previous investigations based on the Lick4227 index have found
Ca in galaxies to be underabundant relative to other $\alpha$ elements,
with the new index, Ca appears to present a behaviour that is more similar to that of other $\alpha$ elements, 
such as Mg. This result alleviates
an otherwise difficult problem, either in nucleosynthesis or in preferential 
depletion onto dust (Thomas \etal 2003), of lowering the Ca abundance
with respect to Mg.

(iv) The difference in Ca abundance from the two index definitions 
lies in a contribution to the blue continuum bandpass from the CN molecule,
coupled with the non-solar abundance ratios in most early-type galaxies. 
Because Ca4227$_r$ uses only a red continuum bandpass, the CN contribution is 
avoided.

(v) We have investigated the effects of uncertainties in wavelength zeropoint, flux calibration, and spectral resolution. For reasonable uncertainties in each, we find  that the effect on the measured Ca4227$_r$ index produces a corresponding uncertainty of up to $\pm$0.05 dex in the derived Ca abundance.

(vi) While the [Ca/Fe] abundances in early-type galaxies, as determined from 
the Ca4227$_r$ index, are now consistent with those of [Mg/Fe],
there remains an unresolved inconsistency between [Ca/Fe] derived from the
Ca I 4227 feature and that determined from the IR Ca triplet (Thomas \etal 2003 and references therein). That problem is reinforced by the recent result of Cenarro \etal (2004) that the Ca abundances inferred from the Ca4227 feature versus those inferred from the Ca II triplet exhibit different trends in cluster versus field galaxies. Clearly, there remain unresolved issues before definitive Ca abundances can be extracted from the integrated spectra of galalxies.

We thank Kristi Concannon and Nelson Caldwell for allowing us to utilize the
CRC galaxy data.  This study was partially funded by NSF grant AST-0406443
to the University of North Carolina. R.P.S. acknowledges financial support from HST Treasurey Program grant GO-09455.05-A to the University of Virginia.

\newpage

\begin{deluxetable}{lll}
\tabletypesize{\small}
\tablewidth{0pt}
\tablenum{1}
\tablecaption{Index Definitions\label{indexdef}}
\tablehead{
  \colhead{Index}&\colhead{Index Passband}&\colhead{Pseudocontinua}}
\startdata
Ca4227$_r$&4221.0-4230.8&4241.0-4257.0\\
Ca4227(Worthey et al. 1994)\tablenotemark{a}&4223.5-4236.0&4212.25-4221.0, 4242.25-4252.25\\
Mg~$b$\tablenotemark{a}&5160.125-5192.625&5142.625-5161.375, 5191.375-5206.375\\
Fe4383\tablenotemark{a}&4370.375-4221.625&4360.375-4371.625, 4444.125-4456.625\\
\enddata
\tablenotetext{a}{Index definitions are from Worthey et al. (1994).}
\label{tab:indexdef}
\end{deluxetable}

\begin{deluxetable}{lll}
\tabletypesize{\small}
\tablewidth{0pt}
\tablenum{2}
\tablecaption{Ca4227$_r$ and Lick Ca4227 Index Measurements (Jones Library Stars)}
\tablehead{
\colhead{Name}& \colhead{Ca4227$_r$}& \colhead{Lick Ca4227}}
\startdata
BD+01\_2341 & -0.0829 & 0.0446 \\
BD+09\_3223 & 0.1229 & 0.1607 \\
BD+11\_2998 & 0.2014 & 0.2702 \\
BD+18\_2890 & 0.2638 & 0.3686 \\
BD+18\_2976 & 0.1453 & 0.1655 \\
BD+25\_1981 & -0.0143 & 0.1327 \\
\enddata
\tablecomments{A portion of Table 2 is shown here for guidance.}
\label{tab:jonesindices}
\end{deluxetable}

\begin{deluxetable}{lllll}
\tabletypesize{\scriptsize}
\tablewidth{0pt}
\tablenum{3}
\tablecaption{Ca4227$_r$ and Lick Ca4227 Index Measurements (Galaxies and 47 Tuc)}
\tablehead{
\colhead{Galaxy ID}& \colhead{Ca4227$_r$}& \colhead{$\pm$1$\sigma$}& \colhead{Lick Ca4227}& \colhead{$\pm$1$\sigma$}}
\startdata
47Tuc & 0.6411 & 0.0099 & 0.5522 & 0.01058  \\
A00368p25 & 1.1209 & 0.117251 & 0.9840 & 0.0942476 \\
A10025 & 1.2044 & 0.0602541 & 1.0070 & 0.0484328 \\
A15572p48 & 0.9717 & 0.117251 & 0.9511 & 0.0942476 \\
A6044p16 & 1.1940 & 0.0618826 & 1.0457 & 0.0497418 \\
A23302p29 & 1.3129 & 0.156335 & 1.0713 & 0.125663 \\
\enddata
\tablecomments{A portion of Table 3 is shown here for guidance.}
\label{tab:galaxyindices}
\end{deluxetable}

\newpage

\clearpage

\end{document}